\documentclass[preprint,amsmath,amssymb,aps,a4,superscriptaddress]{revtex4-2}
\usepackage{graphicx}
\usepackage{pgfplots,gincltex}
\usepackage{txfonts}
\usepackage{hyperref}
\date{\today}
\usepackage[utf8]{inputenc}
\usepackage{amsmath}
\usepackage{braket}

\begin{document}
	\title{	Non-perturbative Generation of Light Antiquark Flavor Asymmetry in Proton}

	\author{Shweta Choudhary}
	\email{shwetaaaru97@gmail.com}
	\affiliation{Department of Physics, Dr. B. R. Ambedkar National Institute of Technology, Jalandhar 144027, India }
	\author{Pranjal Srivastava}
	\email{pranjal.srivastava61@gmail.com}
	\affiliation{Department of Physics, Dr. B. R. Ambedkar National Institute of Technology, Jalandhar 144027, India }
	\author{Harleen Dahiya}
	\email{dahiyah@nitj.ac.in}
	\affiliation{Department of Physics, Dr. B. R. Ambedkar National Institute of Technology, Jalandhar 144027, India }

	\date{\today}
	\begin{abstract}
	We compute the light antiquark flavor asymmetry in the proton using the Chiral Quark Model ($\chi_{\rm QM}$). The distribution functions for the light antiquarks $\bar{d}(x)$ and $\bar{u}(x)$ have been extracted with the help of experimental data from NuSea/E866 and HERMES for the Bjorken$-x$ range $0.015 < x < 0.35$ as well from the most recent SeaQuest data for an extended $x$ range $0.13 < x < 0.45$.  Our results on the $\bar{d}(x)-\bar{u}(x)$,  $\frac{\bar{d}(x)}{\bar{u}(x)}$ and Gottfried Integral $I_G$ are in agreement with the experimental data  and confirm the presence of  enhanced $\bar{d}$ sea whose origin is purely  non-perturbative based on chiral symmetry breaking in QCD.
	\end{abstract}
	\maketitle

\section{Introduction}	
There have been various experimental attempts over the past few decades to measure the flavor asymmetry of sea quarks inside a proton. Initial experiments were based on deep inelastic scattering (DIS) \cite{61201041}, where a high-energy lepton beam collides with the quarks inside a hadron. Since quarks can not exist in free state due to quark confinement, several quark antiquark pairs in hadronized form are produced during collision. This is called fragmentation and fragmentation function ${D_f}^h(z)$ is defined as the probability of producing a hadron $h$ carrying a fraction $z$ of the momentum of the fragmenting quark with flavor $f$ \cite{garvey2001flavor}. These experiments tried to measure quantities like $\bar{d}-\bar{u}$, $\bar{d}/\bar{u}$, $(\bar{d}-\bar{u})/(\bar{u}-\bar{d})$ at different values of the fractional momentum carried by the partons ($x=\frac{Q^2}{2M\nu}$) where $Q^2$ is the square of four momentum transfer carried by virtual photon when a particle interacts with a nucleon during a scattering process, $M$ is the mass of the nucleon and $\nu=E-E^\prime$ is the energy difference between incident and scattered particle \cite{vogt2000physic,klein2016deep}.   DIS can be inclusive or semi-inclusive. In an inclusive DIS experiment, only the scattered lepton is detected and we have no information about the particle that the lepton collided with, but in a semi-inclusive DIS (SIDIS) we can detect or tag the second particle as well, allowing us to distinguish between various quark flavors and  providing explicit quark distributions as a function of $x$.  The HERMES experiment was based on SIDIS \cite{yaschenko2011overview} and it extracted $\bar{d}(x)-\bar{u}(x)$ in the range $0.02<x<0.3$. It used 27.5 GeV positrons which were incident on unpolarized fixed gaseous targets (Hydrogen, Deuterium and Helium-3). The average $Q^2$ value covered by HERMES was $2.3$ GeV$^2$ \cite{miller1999flavor}. The experiment found that $\bar{d}-\bar{u}>0$ over the entire $x$ range and with the help of fragmentation function, $\bar{d}(x)-\bar{u}(x)$ was extracted \cite{duren2000highlights}.

Even though there are multiple ways to investigate flavor asymmetry of sea quarks inside proton, recent attempts prefer experiments based on the Drell-Yan reactions because of their high sensitivity to $\bar{d}/\bar{u}$ ratio \cite{hou2022connected,nagai2022measurements}. In a Drell-Yan reaction, there is annihilation of quark in the hadron beam with the antiquark in the target hadron, resulting in muon pair production. NA51 at CERN was the first such experiment and it reported that the ratio $\bar{d}/\bar{u}=2$ at $x=0.18$ \cite{garvey2001flavor}. Having only a single data point, the dependence of $\bar{d}/\bar{u}$ on $x$ could not be extracted. In order to study the $x-$dependence, NuSea/E866 experiment at Fermilab was conducted to collect the data over a comparatively larger $x$ range with $0.015<x<0.35$ at $Q^2=54$ GeV$^2$. NuSea/E866 data was able to reshape various global fits, however, the experiment was designed for a low $x$ region around $x=0.3$. The experiment reported the ratio $\bar{d}/\bar{u}<1$ \cite{towell2001improved}. In addidtion to $\bar{d}/\bar{u}$ data NuSea/E866 experiment also reported $\bar{d}-\bar{u}$ data which was consistent with the HERMES data. To further investigate the peculiar behavior of  $\bar{d}/\bar{u}$ data of NuSea/E866 around $x=0.3$, SeaQuest/E906 experiment was conducted to specifically collect data in the high $x$ region $0.13<x<0.45$ at $Q^2\approx 29$ GeV$^2$ \cite{dove2021asymmetry,reimer2006opportunities,Dove:2019gyt}. In contrast to its predecessor, the E906 experiment produced a relatively flat dataset that remained above one throughout the experimental range.

The measurement of Gottfried Sum Rule (GSR) was another important evidence to confirm the flavor asymmetry of sea quarks inside the nucleons. It states that if the sea is flavor symmetric then the value of Gottfried integral $I_G=\frac{1}{3} + \frac{2}{3} \int_{0}^{1} [\bar{u}(x)-\bar{d}(x)]dx$ must be $\frac{1}{3}$. However, the GSR has been observed to be violated in all the previously conducted experiments. Its violation is a clear indication of asymmetric sea and role of non-perturbative effects inside the proton. All of the experiments have found the value of $I_G$ to be less than $\frac{1}{3}$. The data from various experiments has been listed in Table \ref{GSR}. 
	
	\begin{center}
		\begin{table}[h!]
			\begin{tabular} { | p {5 cm}| p {8 cm} | }
				\hline
		
				Experiment &  \hspace{1.5 cm} Gottfried Integral $I_G$  \\
				\hline
				SLAC & \hspace{2 cm}$0.200\pm0.040$  \\
				EMC &  \hspace{2 cm}$0.235\pm0.099$  \\ 
				NMC &  \hspace{2 cm}$0.266\pm0.005$  \\
				HERMES &  \hspace{2 cm}$0.226\pm0.020$ \\ 
				NuSea/E866 &  \hspace{2 cm}$0.254\pm0.005$\\   
				\hline
			\end{tabular}
			\caption{\label{GSR}Experimentally measured values of Gottfried integral.}
		\end{table}
	\end{center} 
	
There have been various theoretical ideas to explain the measured flavor asymmetry of sea quarks inside the proton. It was posited by Field and Feynman that the gluons annihilate into quark antiquark pairs inside the proton with a valence structure $uud$. Since the creation of $u\bar{u}$ pairs is suppressed compared to $d\bar{d}$ pairs due to the Pauli's exclusion principle, there is an excess of $\bar{d}$ over $\bar{u}$ \cite{song2011light}. It is however now known that this perturbative phenomenon called Pauli blocking contributes less than $1\%$. Therefore, to successfully explain the flavor asymmetry in proton, there is a need to adopt non-perturbative models. Two of the most significant non-perturbative models that can explain flavor asymmetry inside the proton are (i) Pion Cloud Model (PCM) and (ii) Chiral Quark Model $(\chi_{\rm QM})$.

In the PCM \cite{christiansen2001light,fries,sanchis2014pion,avila2003pion}, the flavor asymmetry arises due to the emission of a virtual pion by the proton and the transitory state consists of a pion and a baryon. The state of proton in PCM can be expressed as $\ket{p}=(1-a-b)\ket{p_0}+a\ket{n\pi^+}+b\ket{\Delta^{++}\pi^-}$ with $\ket{p_0}$ being the proton state with assumed symmetric sea. The asymmetry in this case can thus be attributed to the existence of virtual pions in other available states \cite{melnitchouk1998dynamics}. Here, $a$ and $b$ represent the transition probability amplitudes of the proton going into one of the available states consisting of $n\pi^+$ or $\Delta^{++}\pi^-$. According to Sullivan mechanism \cite{song2011light}, since the number of valence $u$ quarks in a proton is greater than the number of valence $d$ quarks, the state $p(uud)\rightarrow n(udd)\pi^+(\bar{d}u)\rightarrow p(uud)$ is more likely to occur as compared to the state  $p(uud)\rightarrow \Delta^{++}(uuu)\pi^-(\bar{u}d)\rightarrow p(uud)$ \cite{dahiya2021quark}. In addition, because of the large mass of $\Delta$ resonance, $n\pi^+$ state is preferred over the $\Delta^{++}\pi^-$ which results in excess of $\bar{d}$ over $\bar{u}$ inside the proton \cite{brodsky1996quark}. Similar to this is another non-perturbative approach which explains the flavor asymmetry of sea quarks in the instanton model and is based on the interaction between a quark and an instanton which is basically a fluctuation of gauge field. This interaction produces a quark antiquark pair of different flavor than the original quark whilst flipping the helicity of original quark. Therefore, this model can inherently tackle the proton spin puzzle along with proton's flavor structure but it somehow over predicts the $\bar{d}/\bar{u}$ ratio when compared to experiments giving $\bar{d}/\bar{u}=4$ as $x\rightarrow1$ \cite{isenhower2001proposal}.

On the other hand, $\chi_{\rm QM}$ \cite{dahiya2001chiral,thomas2000flavor} also employs virtual mesons to explain the sea quark flavor asymmetry but in this model, the  virtual mesons called Goldstone Bosons (GBs) ($\pi,K,\eta$) are emitted by the valence quarks inside the proton, whereas in PCM,  the virtual pions are emitted by the proton itself. Apart from these, another model based on quark-quark GB interaction is the GBE model \cite{GBE} which has been successful in fitting the baryon spectrum \cite{GBE-1998}, predicting the electromagnetic $G_E$, $G_M$ form factors \cite{GBE-2001}, axial $G_A$ and pseudoscalar $G_P$ form factors \cite{GBE-2002}. In $\chi_{\rm QM}$, the valence $u$ and $d$ quarks can momentarily exist as $u\rightarrow d\pi^+$ and $d\rightarrow u\pi^-$ respectively \cite{ohlsson1997properties,dahiya2007x}. Since there are more valence $u$ quarks inside a proton, more $\pi^+=\bar{d}u$ will be produced compared to $\pi^-=d \bar{u}$, resulting in an excess of $\bar{d}$ over $\bar{u}$. The $\chi_{\rm QM}$ has been  successful in explaining the proton spin puzzle \cite{dahiya2022flavor},  magnetic moment associated with the octet, decuplet baryons,  decuplet and octet baryon resonances \cite{sharma2010spin,dahiya2003octet} and unlike the PCM, it works in the interior of nucleon \cite{sharma2010quark,dahiya2017octet}. Keeping in mind the successes of $\chi_{\rm QM}$ for a wide range of observables, we attempt to extract the sea quark distribution functions $\bar{d}(x)$ and $\bar{u}(x)$ as a function of $x$ inside the proton. Using these distribution functions we compute the asymmetries like $\bar{d}(x)-\bar{u}(x)$,  $\frac{\bar{d}(x)}{\bar{u}(x)}$ and Gottfried Integral $I_G$ and compare them with the  experimental data from NuSea/E866 and HERMES for  range $0.015 < x < 0.35$ as well from the most recent SeaQuest data for  $0.13 < x < 0.45$.

\section{Chiral Quark Model ($\chi_{\rm QM}$)}
It has been well established that the models based on perturbative QCD are unable to explain the sea quark asymmetry \cite{song2011light} and they either have a small effect or the results are inconsistent with the experiments. Moreover, in the low energy region, perturbative QCD cannot be used because the value of running coupling constant becomes large and higher order terms in the perturbation can no longer be neglected. As a result, non-perturbative phenomenons like chiral symmetry breaking and quark confinement become prevalent \cite{histcqm}. The $\chi_{\rm QM}$ is the most appropriate model to study the flavor asymmetry and to find the $x$ dependence of light sea quarks inside a proton since it functions in the energy range of 300 MeV to 1 GeV \cite{ohlsson1997properties} where non-perturbative QCD is applicable. In $\chi_{\rm QM}$, the GBs which include $\pi$, $K$, $\eta$ and $\eta'$, are created from the valence quark transitions to the GBs  where a valence quark $q$ makes the transition into a GB and quark of some other flavor $q^{\prime\mp}$ \cite{dahiya2007x}. If the probability of such a transition is given by a parameter $\varepsilon$ then the transition probabilities of a quark going into $K$, $\eta$ and $\eta^\prime$ are given by $\varepsilon\lambda^2$, $\varepsilon\kappa^2$ and $\varepsilon\xi^2$ respectively where $\lambda$, $\kappa$ and $\xi$ are the probability parameters. The mass is provided to these GBs through spontaneous chiral symmetry breaking \cite{scherer2003introduction,dahiya2011strangeness}. 

The interaction Lagrangian between a quark and GB at the lowest order is represented by $\mathcal{L}_I=g_{8} \bar{\psi} \varphi^{\prime} \psi$ \cite{dahiya2007x,ding2005nucleon}, where $g_{8}$ is a coupling constant whose modulus squared $|g_{8}|^{2}\propto\varepsilon$ gives the probability of transition of a quark into a GB which further decays into a quark antiquark pair i.e $q\rightarrow q^{\prime}+GB\rightarrow q^{\prime} + q\bar{q^{\prime}}$. $\bar{\psi}$ and $\psi$ are the Dirac antiquark and quark fields respectively and $\varphi^{\prime}$ is a $3\times3$ matrix representing nonet of GBs. The interaction term $\mathcal{L}$ can be expressed in terms of the quark and GB fields as
\newline
$$\mathcal{L}_I=g_{8}\begin{pmatrix}
	\bar{u} & \bar{d} & \bar{s}  
\end{pmatrix}\left(\begin{array}{ccc}\frac{\pi^{0}}{\sqrt{2}}+\kappa \frac{\eta}{\sqrt{6}}+\frac{\xi \eta^{\prime}}{\sqrt{3}} & \pi^{+} & \lambda K^{+} \\ \pi^{-} & -\frac{\pi^{0}}{\sqrt{2}}+\frac{\kappa \eta}{\sqrt{6}}+\frac{\xi \eta^{\prime}}{\sqrt{3}} & \lambda K^{0} \\ \lambda K^{-} & \lambda \overline{K^{0}} & -\frac{\kappa 2 \eta}{\sqrt{6}}+\frac{\xi \eta^{\prime}}{\sqrt{3}}\end{array}\right)
\left(\begin{array}{l}u \\ d \\ s\end{array}\right).\\
$$\\
On solving the above equation after substituting the mesons with their corresponding flavor wavefunctions, we obtain the flavor wavefunctions $\psi(u)$, $\psi(d)$ and $\psi(s)$ corresponding to the $u, d$ and $s$ quarks respectively \cite{dahiya2015quark}. For the case of $u$ quark we have
\begin{align}
	\psi(u) &\approx  \left(\frac{1}{2}+\frac{\kappa}{6}+\frac{\xi}{3}\right)u(u\bar{u})+\left(-\frac{1}{2}+\frac{\kappa}{6}+\frac{\xi}{3}\right)u(d\bar{d}) +\left(-\frac{\kappa}{3}+\frac{\xi}{3}\right)u(s\bar{s})+(u\bar{d})d+\lambda(u\bar{s})s.
\end{align}
Since $|\psi(u)|^{2}$ gives the probability of a ``$u$ quark'' to emit a GB, the coefficients associated with $q=u,d,s$ and $\bar{q}=\bar{u},\bar{d},\bar{s}$ give the creation probability of $q$ and $\bar{q}$ when a GB is emitted by a ``$u$ quark'' \cite{ohlsson1997properties}. We have
	\begin{align}	
		|\psi(u)|^{2}&=\varepsilon\left[\frac{7}{4}+\frac{\kappa}{6}+\frac{\xi}{3}+\frac{\kappa \xi}{9}+\lambda^{2}+\frac{7 \kappa^{2}}{36}+\frac{4 \xi^{2}}{9}\right] u+\varepsilon\left[\frac{1}{4}+\frac{\kappa}{6}+\frac{\xi}{3}+\frac{\kappa \xi}{9}+\frac{\kappa^{2}}{36}+\frac{\xi^{2}}{9}\right] \bar{u}\nonumber\\
		&+\varepsilon\left[\frac{5}{4}-\frac{\kappa}{6}-\frac{\xi}{3}+\frac{\kappa \xi}{9}+\frac{\kappa^{2}}{36}+\frac{\xi^{2}}{9}\right](d+\bar{d})+\varepsilon\left[-\frac{2 \kappa \xi}{9}+\lambda^{2}+\frac{\kappa^{2}}{9}+\frac{\xi^{2}}{9}\right](s+\bar{s}).
	\end{align}
Similar expressions can be written for $|\psi(d)|^{2}$ and $|\psi(s)|^{2}$. The total probability ($P_{q\rightarrow GB}$) of GB emission by a quark $q$  can be obtained by adding $|\psi(u)|^{2}$, $|\psi(d)|^{2}$ and $|\psi(s)|^{2}$ and we get 
	\begin{align}
		P_{q\rightarrow GB}&= 2\varepsilon\left(\frac{3}{2}+\lambda^{2}+\frac{\kappa^{2}}{6}+\frac{\xi^{2}}{3}\right) +\varepsilon\left(2 \lambda^{2}+\frac{2 \kappa^{2}}{3}+\frac{\xi^{2}}{3}\right).
	\end{align}
Here the first term comes from $|\psi(u)|^{2}+|\psi(d)|^{2}$ and the second term from $|\psi(s)|^{2}$. Simplifying further we have
\begin{align}
	P_{q\rightarrow GB}&= \varepsilon\left(3+4 \lambda^{2}+\kappa^{2}+\xi^{2}\right).
\end{align}
If the total probability of GB emission by a quark q is $P_{q\rightarrow GB}$ then the probability of no GB emission by the quarks is $P_{q\not\rightarrow GB}=P_{q}=1-P_{q\rightarrow GB}$. The probabilities that a $u$, $d$ or $s$ quark does not emit a GB are given as
	\begin{align}
		P_{u}=P_{d}&=1-\varepsilon\left(\frac{3}{2}+\lambda^{2}+\frac{\kappa^{2}}{6}+\frac{\xi^{2}}{3}\right),\nonumber\\
		P_{s}&=1-\varepsilon\left(2 \lambda^{2}+\frac{2 \kappa^{2}}{3}+\frac{\xi^{2}}{3}\right).
	\end{align}

In the $\chi_{\rm QM}$, we can now represent a GB emission by a quark as 
\begin{equation}
	q \rightarrow P_{q} q+ |\psi(q)|^{2}.
\end{equation}
Since the flavor content of proton is $p_{0}=2u+ d$,  we can modify $p_{0}$ to $p_{1}=2 P_{u} u+2 |\psi(u)|^{2}+P_{d} d+|\psi(d)|^{2}$ after considering the GB emission by quarks. On expanding $p_{1}$ by substituting the relevant expressions, collecting the coefficients of $\bar{u}$ and $\bar{d}$ explicitly and then subtracting them gives us the difference 
	
	\begin{align}
		\bar{u}-\bar{d}&= \varepsilon\left(\frac{7}{4}+\frac{\kappa^{2}}{12}+\frac{\xi}{3}+\frac{\xi^{2}}{3}+\frac{\kappa}{6}+\frac{\kappa \xi}{3}\right)-\varepsilon\left(\frac{11}{4}+\frac{\kappa^{2}}{12}-\frac{\xi}{3}+\frac{\xi^{2}}{3}-\frac{\kappa}{6}+\frac{\kappa \xi}{3}\right)
	\end{align}
Similarly, by dividing the coefficients of $\bar{u}$ and $\bar{d}$ we obtain the ratio $\bar{u}/\bar{d}$ and is expressed as 
\begin{equation}
	\frac{\bar{u}}{\bar{d}}=\dfrac{21+\kappa^{2}+4\xi+4\xi^{2}+2\kappa+4\kappa \xi}{33+\kappa^{2}-4\xi+4\xi^{2}-2\kappa+4\kappa \xi}.
\end{equation}
By substituting the coefficients in the above equations we get a measure of asymmetry of light sea quarks in proton. To get the measure of asymmetry from experimental data we calculate a quantity called Gottfried Integral $I_G$.
\begin{align}
	I_G(x_{\rm min},x_{\rm max})&= \int_{x_{\rm min}}^{x_{\rm max}} \dfrac{F_{2}^{p}(x)-F_{2}^{n}(x)}{x} dx =\frac{1}{3} + \frac{2}{3} \int_{x_{\rm min}}^{x_{\rm max}} [\bar{u}(x)-\bar{d}(x)]dx,
\end{align}
where $F_{2}^{p}(x)$ and $F_{2}^{n}(x)$ are proton and neutron structure functions and $(x_{\rm max}-x_{\rm min})$ is the experimental range. Measurements in the unmeasured $x$ regions will have the possibility to impose significant constraints in different $x$ regions  for GSR \cite{garvey2001flavor}. According to GSR, if the proton is only composed of valence quarks ($\bar{u}(x)=\bar{d}(x)=0$) or if the proton sea is symmetric ($\bar{u}(x)=\bar{d}(x)$) then the integral $I_G$ must be $\frac{1}{3}$. A value other than $\frac{1}{3}$ would indicate asymmetry of light sea quarks. All the experimental collaborations so far have reported a value of $I_G<1/3$ which can be interpreted as $\bar{d}(x)>\bar{u}(x)$ implying an excess of down antiquarks in the sea. Using extracted distribution functions $\bar{d}(x)$ and $\bar{u}(x)$ we have calculated the value of $I_G$. 
	
\section{Results and discussion}
The input parameters of the coefficients used in the present work have been taken to be $\varepsilon=0.114\pm 0.01$, $\kappa=0.45\pm 0.04$, $\xi=-0.75 \pm 0.07$ following Ref. \cite{song2001quark}. The results for the case of 	$\bar{u} - \bar{d}$ and $\bar{u}/\bar{d}$  obtained are 
\begin{eqnarray}
	\bar{u} - \bar{d}&=&-0.118 \pm 0.015, \nonumber\\
	\bar{u}/\bar{d}&=&0.652 \pm 0.07. 
\end{eqnarray}
It is confirmed from these values that, according to $\chi_{\rm QM}$, there is indeed a flavor asymmetry of light sea antiquarks and also that $\bar{d}>\bar{u}$ in proton. This can further be compared with the experimental data by first fitting the  distribution functions over the Bjorken variable and then using these distribution functions for the computation of the flavor asymmetries.

In order to fit the  light quark distribution functions $\bar{d}(x)$ and $\bar{u}(x)$ in proton we begin with a simple form of PDF  $(x)^A(1-x)^B$ as an ansatz \cite{article,lai2000global,alekhin2005parton,dahiya2020parton}. In the low $x$ region $(x)^A$ term dominates and in the large $x$ region $(1-x)^B$ term dominates \cite{melnitchouk1998dynamics,accardi2020shape,jacobE}. However, this form of function as it is, does not provide the best fit for the dataset. We thus multiply our initial guess with a polynomial $P(x)=(1+Cx)$ which is responsible for shape of the function in the middle $x$ region. The functional form finally used for the distribution functions is taken to be $\bar{q}(x)=(Nx)^A(1-x)^B(1+Cx)$. By fitting the distribution function to $\bar{d}/\bar{u}$ dataset from NuSea and SeaQuest, we extract $\bar{d}(x)$ and $\bar{u}(x)$ using the values of the obtained parameters $N$, $A$, $B$ and $C$ listed in Table \ref{parameters-fitted}. 

\begin{center}
	\begin{table}[h!]
		\begin{tabular} { | c| c |c | c | c |c | }
	
			\hline
			& ~$\chi_{QM}$ & ~$N$ & ~$A$ & ~$B$ & ~$C$ \\
			\hline
			$\bar{d}(x)$& 0.3391$\pm 0.04$ & $1.50\pm0.15$ & $-1.000\pm0.100$  & $9.00\pm0.90$ & $8.0000\pm0.8000$ \\
			$\bar{u}(x)$& 0.2211$\pm 0.02$ & $1.00\pm0.10$ & $-0.999\pm0.099$  & $7.50\pm0.85$  & $1.0875\pm0.1088$\\
			
			\hline
		\end{tabular}
		\caption{\label{parameters-fitted} Parameter values for the fitted function $\bar{q}(x)=(Nx)^A(1-x)^B(1+Cx)$ representing antiquark distribution inside a proton.}
	\end{table}
\end{center}

We would like to mention here that to determine the proximity of the data points to the fitted curve, we have calculated $\frac{\chi^2}{N_{pts}}$, where $N_{pts}$ is the total number of data points in the experiment and $\chi^2$ is chi-squared value. For the SeaQuest experiment, $\frac{\chi^2}{N_{pts}}$ is $1.70$ and for the NuSea experiment, it is $2.63$. This makes our model, with the parameters given in Table II, in better agreement with the SeaQuest data as compared to the NuSea data. This is also partially due to the NuSea datapoint at $x=0.315$ which if excluded gives an improved value of $\frac{\chi^2}{N_{pts}}$ as $2.18$.

%

	\begin{figure}[h!]
		\centering
	\includegraphics[width=14cm]{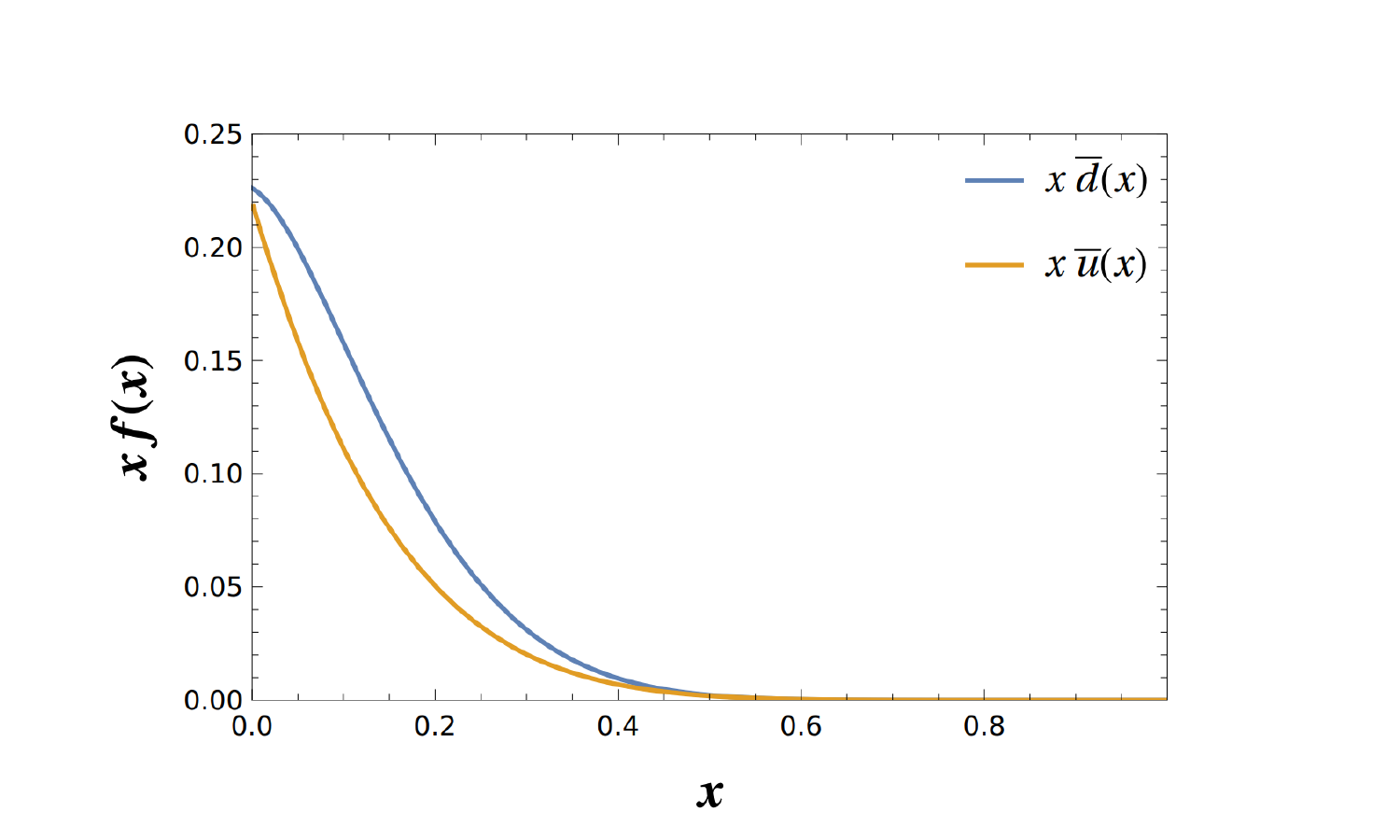}
		\caption{ Light sea quark  distributions inside a proton as a function of Bjorken$-x$.}
		\label{xfx}
	\end{figure}

	\begin{figure}[h!]
	\centering
	\includegraphics[width=14cm]{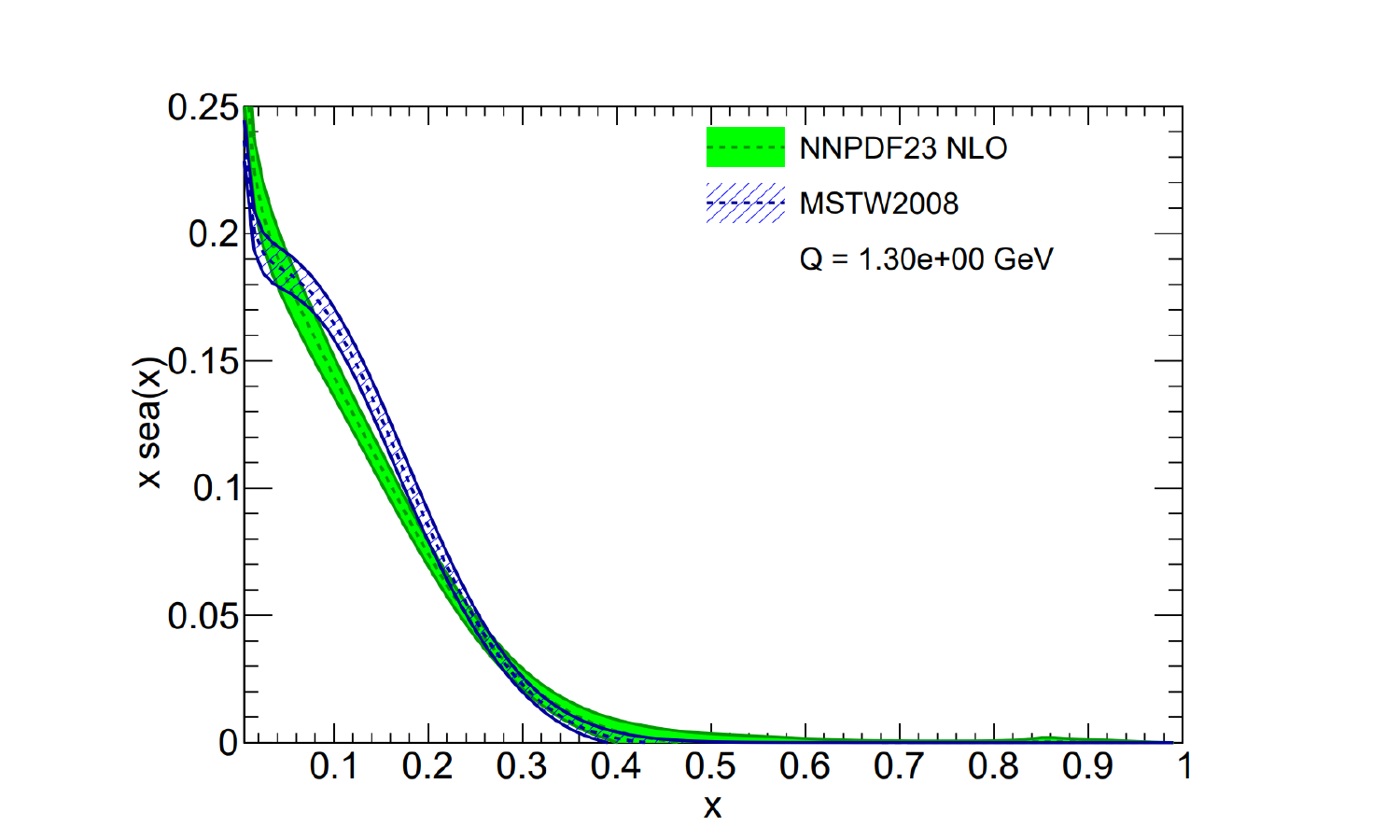}
	\caption{ Light sea quark distributions inside a proton as a function of Bjorken$-x$ obtained from NNPDF23 NLO  and MSTW2008.}
	\label{xfx-comparison}
\end{figure}
	
In order to study the distribution of the sea quarks ($\bar{q}(x)$) in the entire Bjorken range, we plot $xf(x)$ for proton in Fig. \ref{xfx}. Here $f(x)=\bar{q}(x)$ with $q=u,d$. In order to compare our results with the results from other parametrizations of light antiquark distributions, we have also presented the results from NNPDF23 NLO \cite{nnpdf} and MSTW2008 \cite{MSTW2008} in Fig. \ref{xfx-comparison}. Since the natural scale of the $\chi$QM is $0.3 < Q^2 < 1$ GeV$^2$, we have  taken the results of these parametrizations at Q=1.3 GeV.  As expected,  we observe that with its peak near the origin, the sea quark distributions decay rapidly and almost vanish around $x=0.35$.  From these plots, some general aspects of the sea quark distributions can be described. The  sea quark or the antiquark distributions underscore that $\bar{d}(x)>\bar{u}(x)$ only in the sea quark dominated region or the lower $x$ region. As the value of $x$ increases, the sea contributions decrease and for $x > 0.35$, the sea quark asymmetry vanishes. This drop is very rapid which is in line with the experimental observations of  DIS experiments \cite{hera,chekelian2016proton}. The  sea quark distributions in Fig. \ref{xfx} are also in line with the antiquark distributions obtained by the other well known parametrizations \cite{MMHT14,MSHT20,CT18,HERAPDF2.0} as well.

	\begin{figure}[h!]
		\centering
		\includegraphics[width=14cm]{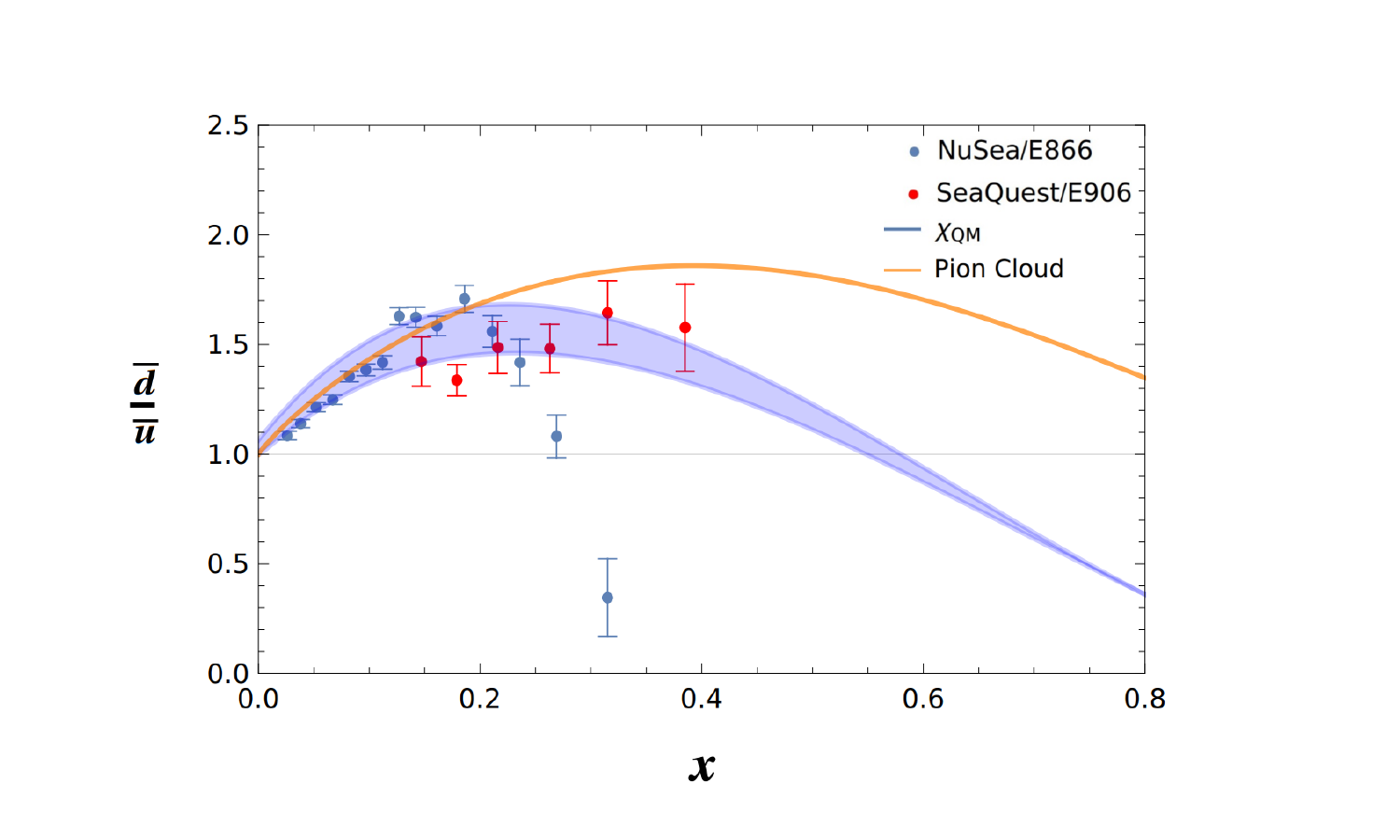}
		\caption{ The  $\chi_{\rm QM}$ results for $\bar{d}(x)/\bar{u}(x)$ in comparison to the pion cloud model. The data from  NuSea/E866 and SeaQuest/E906 has also been included.
		}
		\label{dbaroverubar}
	\end{figure}
	
After discussing the qualitative bahavior of the sea quark asymmetry, we now discuss the asymmetry measured through the $\bar{d}/\bar{u}$  quantitatively. We show in Fig. \ref{dbaroverubar} the ratio $\bar{d}(x)/\bar{u}(x)$ as obtained in the $\chi_{\rm QM}$. For the sake of comparison we have also shown the results from PCM \cite{alberg2019chiral}. The experimental data for  $\bar{d}(x)/\bar{u}(x)$  has also been shown in the same plot from the SeaQuest/E906 \cite{dove2021asymmetry} and NuSea/E866 \cite{towell2001improved,zhu2016determination} experiments. The asymmetry based on $\chi_{\rm QM}$ in Fig. \ref{dbaroverubar} shows that the ratio $\bar{d}(x)/\bar{u}(x)$ lies between 1 and 1.57 in the range $0<x<0.57$ with a peak at around $x=0.229$. Beyond this $x$ value, the function gradually decays, reaches unity at $x=0.57$ and thereafter drops further. The $\chi_{\rm QM}$ results, on the one hand, are consistent to some extent with the NuSea/E866 data  \cite{towell2001improved,zhu2016determination}  showing a downward trend for slightly higher values of $x$. This, on the other hand, is different from the flat trend of the SeaQuest/E906 data \cite{dove2021asymmetry}. It would be important to mention here that these experiments have been performed at different $Q^2$ values so there is no consensus regarding the agreement of data with each other and with other theoretical models. The theoretical studies also take different assumptions and model parameters which makes it difficult to make explicit correspondence at each $x$ value. The PCM explains the NuSea/E866 data for $x<0.20$ but somehow fails to explain the downward trend of $\bar{d}/\bar{u}$ for $x>0.2$ \cite{alberg2019chiral}. In the PCM the ratio $\bar{d}/\bar{u}$ lies between 1 and 1.5 in range $0<x<0.15$, between 1.5 and 2 in range $0.15<x<0.4$ with a peak at $x=0.4$. Beyond this $x$ value the ratio decays to unity but very steadily. The ratio $\bar{d}(x)/\bar{u}(x)$ in PCM initially rises because, with increasing $x$, the contribution of $n\pi^+$ increases upto $x=0.4$ resulting in an increment in $\bar{d}$. However, for $x>0.4$  contributions of $\Delta^{++}\pi^-$ become more pertinent, introducing more and more $\bar{u}$ which causes the ratio to decrease \cite{vogt2000physic,sharma2010quark,dahiya2011nonperturbative}. We will now attempt to compare our results at a specific value of $x$. The  SeaQuest/E906 data for different $x$ values \cite{e906Reimer} is in fair agreement with the $\chi_{\rm QM}$ values   for $\bar{d}(x)/\bar{u}(x)$ at the points of $x$ for which data is available.   The results become slightly different at higher values of $x$ which is again because of different initial conditions and parameters. This can perhaps be substantiated further by future measurements in the entire $x$ region which would have important implications for the subtle features of the quark sea asymmetry. 

	\begin{figure}[h!]
		\centering
		\includegraphics[width=14cm]{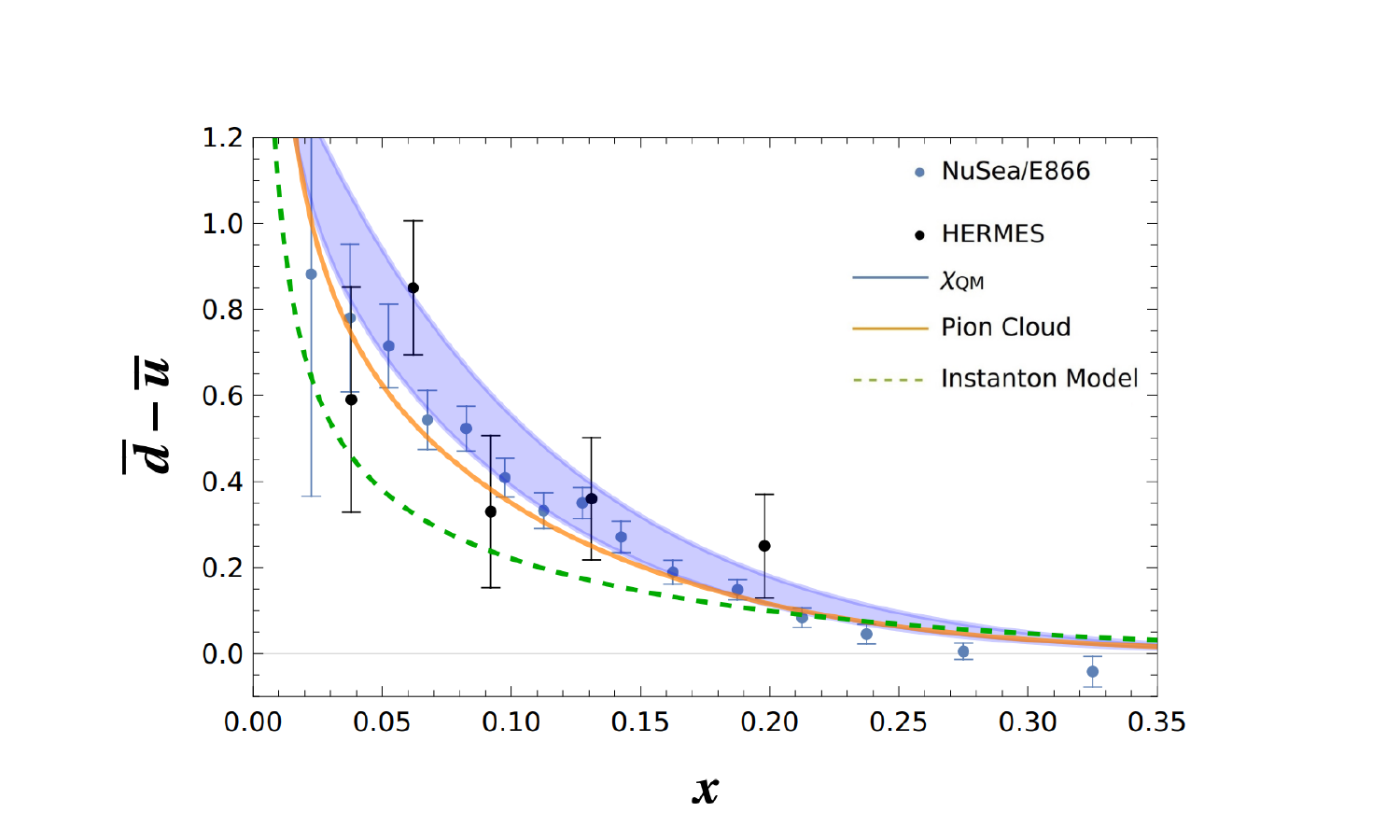}
		\caption{The  $\chi_{\rm QM}$ results for $\bar{d}(x)-\bar{u}(x)$. Pion cloud model and instanton model results have also been presented and compared with the HERMES and NuSea/E866 data.}
		\label{dbar-ubar}
	\end{figure}
	
Further, we can plot the asymmetry $\bar{d}(x)-\bar{u}(x)$ using our distribution functions and compare it with other non-perturbative models like PCM and instanton model.  It is important to mention here that this asymmetry is obtained in independent experiments and we have data for  $\bar{d}(x)-\bar{u}(x)$ from NuSea/E866 \cite{towell2001improved} and HERMES \cite{miller1999flavor}.  SeaQuest does not provide data on $\bar{d}(x)-\bar{u}(x)$. The results alongwith available data 
have been presented in Fig. \ref{dbar-ubar}. We observe that the distribution function $\bar{d}(x)-\bar{u}(x)$ is in reasonably  good agreement with both the data sets available. In the case of NuSea data, the experimental value of $\bar{d}-\bar{u}=0.259\pm0.04$ at $x=0.142$ matches with our results which is $0.286 \pm 0.02$. Similarly, HERMES experimentally reports $\bar{d}-\bar{u}=0.36\pm0.14$ at $x=0.131$ and our value is $0.327 \pm 0.03$ which is also within the error limits. The distribution function based on $\chi_{\rm QM}$ does not become negative for any value of $x$ in the experimental range $0.015<x<0.35$ and shows a significant excess of $\bar{d}$ over $\bar{u}$ for low-$x$ values but for $x>0.3$ the difference $\bar{d}(x)-\bar{u}(x) \rightarrow 0$ implying an absence of sea quarks in this region \cite{dahiya2014asymmetries}.  While the PCM results  are in good agreement with the data, both the $\chi_{\rm QM}$ and instanton model \cite{towell2001improved} show a little deviation. Qualitatively, all the models clearly accentuate the importance of mesons in explaining the flavor asymmetry of sea quarks. The degree of numerical agreement in each case can be assessed only after the data gets
refined.

As discussed earlier, the violation of Gottfried Sum Rule i.e $I_G<\frac{1}{3}$ implies $\bar{d}>\bar{u}$ \cite{amaudruz1991gottfried,dahiya2014asymmetries}. For our distribution function calculated in $\chi_{\rm QM}$, $I_G(x_{\rm min},x_{\rm max})$ comes out to be $0.263 \pm 0.02$ in the range $0.015<x<0.45$. This is comparable to the results of NMC \cite{shao2010sea} and NuSea/E866 \cite{towell2001improved}. On extrapolating the integral by including values from unmeasured high and low $x$ regions where no experimental data is available, we obtain $I_G(0,1)=0.219 \pm 0.02$ which is still less than $\frac{1}{3}$ and consistent with previously measured experimental values. The value however, is lower for the extrapolated $x$ region and is consistent with the SeaQuest data.  This is mainly because the  $\bar u$ and $\bar d$ difference is largely negative at very small $x$ and negligible at large $x$.

\section{Summary and conclusion}

Two successful experimental processes which measure the $x$ dependence of light sea quarks $\bar{d}$ and $\bar{u}$ are the semi-inclusive deep inelastic scattering (SIDIS) and Drell-Yan processes. HERMES experiment reported $\bar{d}(x)-\bar{u}(x)$ in the range $0.02<x<0.3$ by employing the former technique whereas NuSea/E866 reported $\bar{d}(x)-\bar{u}(x)$ and $\bar{d}(x)/\bar{u}(x)$ in the range $0.015<x<0.35$ using the latter. Drell-Yan process is particularly sensitive to the $\bar{d}(x)/\bar{u}(x)$ measurements, therefore, SeaQuest/E906 experiment (a successor of NuSea/E866 experiment) was conducted to measure the $\bar{d}(x)/\bar{u}(x)$ ratio in a slightly higher and wider range of $0.13<x<0.45$.  Since these experiments provide data from moderately low to valence-like and high-$x$ region, they form an excellent testing ground for non-perturbative QCD models which try to explain flavor asymmetry of sea quarks inside a proton. It would be important to mention here that the different experiments have been performed at different $Q^2$ values so the results cannot be compared quantitatively at each value of $x$ but only qualitatively for the available $x$ range. Chiral Quark Model ($\chi_{\rm QM}$) reasonably  explains the proton spin problem  and many other physical phenomena related to baryons, therefore, it becomes a natural choice for explaining the flavor asymmetry between the $\bar{u}$ and $\bar{d}$ in proton through the generation of Goldstone bosons by the valence quarks in the interior of proton. In order to study the variation of these asymmetries in the entire $x$ range, we extracted the quark distribution functions from the NuSea/E866 and SeaQuest/E906 data and then plotted the asymmetries w.r.t $x$.  The results confirm the flavor asymmetry of light antiquarks and more precisely an excess of $\bar{d}$ compared to $\bar{u}$ inside the proton. We also observe that the sea quarks dominate at the lower values of $x$ but their contribution vanishes for higher values of $x$ where the valence quarks dominate. The results are  also compared with other non-perturbative models like pion cloud and instanton model. The value of Gottfried Integral $I_G$ is also calculated in this work and is found to be in good agreement with the experimental values from NuSea/E866 and NMC in the measured $x$ region. In the extrapolated $x$ region,  future Drell-Yan experiments at large $x$ would not only provide further constraints  to understand the origin of quark sea asymmetry but also confirm the not so well understood non-perturbative component in obtaining the asymmetry.

\section{Data Availability}

This manuscript has no associated data.

\section{Acknowledgement}
H.D. would like to thank the Science and Engineering Research Board, Department of Science and Technology, Government of India through the grant (Ref No. MTR/2019/000003) under MATRICS scheme for financial support.

\end{document}